\documentclass[english,a4paper,12pt]{article}
\usepackage{graphicx}
\newcommand{\rytau}{RY~Tau\ }
\newcommand{\cii}{[C~{\sc II}]\ }
\newcommand{\ciii}{C~{\sc III}]\ }

\newcommand{\siii}{Si~{\sc III}]\ }
\newcommand{\civ}{C~{\sc IV}\ }

\newcommand{\oxigen}{[O~{\sc II}]\ }
\newcommand{\mgii}{Mg~{\sc II}\ }
\newcommand{\mgi}{Mg~{\sc I}\ }

\begin{document}

\begin{titlepage}

\begin{center}
\textbf{\large Evidence for stellar driven outflows\\
from the Classical T Tauri Star RY Tau }
\end{center}
\vskip 1cm
\begin{center}
\textbf{Ana I. G\'omez de Castro$^{1}$ and Eva Verdugo$^{2}$}
\end{center}
\begin{tabular}{ll}
$^{1}$ & Instituto de Astronom\'{\i}a y Geodesia (CSIC-UCM)\\
       & Facultad de CC. Matem\'aticas, Univ. Complutense \\
       & Madrid, E28040 Spain\\
$^{2}$ & European Space Astronomy Centre (ESAC) \\
       & Research \& Scientific Support Dept.- ESA \\
       & P.O.Box 50727, Madrid, E28080 Spain \\                                 \\
\end{tabular}
\begin{center}
{\bf ABSTRACT} \vskip 0.5cm
\end{center}
RY~Tau is a rapidly rotating Classical T Tauri star observed close
to edge-on. The combination of new HST/STIS observations obtained
in 2001 with HST/GHRS Archive data from 1993 has allowed us to
get, for the first time, information on the thermal structure and
the velocity law of the wind. The repeated observations of the
\siii\ and \ciii\ lines show a lack of changes with time in the
blue side of the profile (dominated by the wind contribution).
Very high temperature plasma ($\log T_e = 4.8$) is detected at
densities of $9.5 \leq \log n_e {\rm (cm^{-3})}\leq 10.2$
associated with the wind. The emitting volumes are $\sim
(0.35R_{\odot})^3$ suggesting a stellar origin. The wind
kinematics derived from the profiles (\siii, \ciii and \oxigen)
does not satisfy the theoretical predictions of MHD centrifugally
driven disk winds. The profiles' asymmetry, large velocity
dispersions and small variability as well as the small emitting
volumes are best explained if the wind is produced by the
contribution of several outflows from atmospheric open field
structures as those observed in the Sun.


\end{titlepage}

\section{Introduction}

Winds from T Tauri Stars (TTSs) were assumed to be cold,
in spite of the observational evidence of our own Sun, because
they were first detected as bipolar molecular outflows (Loren et al 1981)
and optical jets with low ionization fractions.
The detection of \ciii and \siii semiforbidden line emission
from \rytau and RU~Lup wind (G\'omez de Castro
\& Verdugo, 2001; hereafter GdCV) provided the first evidence
that electron temperatures as
high as  $\log Te(K) \simeq  4.5$ could be achieved at the base of
the TTSs outflows; this high
temperature was derived from ultraviolet (UV) observations and
could not be inferred from the classical optical/infrared/radio tracers.
Recently, Dupree et al (2005) have claimed that O~{\sc VI}
absorption has been detected in TW~Hya and T~Tau.
These new data represent a challenge to the current theories
for outflow generation in TTSs.
In this letter, we provide observational
evidence of hot winds being driven from the stellar atmosphere
of RY~Tau, a rapidly rotating TTS ($v\sin i = 51.6$ km/s, Hartmann
\& Stauffer, 1989). RY Tau has been classified as an UX~Ori
star: a pre-main sequence star which shows
aperiodic eclipse-like minima caused by
variable obscuration produced by circumstellar dust associated with the disk
(Grinin, 1992). This suggests that the disk is seen close to edge-on.
Also,  the high projected rotation velocity of
the stellar atmosphere suggests that the star inclination is
close to 90$^o$.

\section{Data}

Some few spectral tracers have been selected for the subsequent
study, namely, the \civ, \ciii, \cii, \siii and \oxigen
lines observed in the UV spectrum of RY~Tau
(see Table~1 for more details on the observations and the
line fluxes, and
Fig.~1 for the profiles). All data are public and can
be withdrawn from the HST Archive (they were obtained either
by us or by other HST users). Data have been processed with the
Routine Science Data Pipeline (RSDP). The major source of
inaccuracy in the calibration is the centering
of the target in the aperture that can account for as much as
20~km~s$^{-1}$ in the wavelength zeropoint accuracy of the GHRS data
(see GdCV) and $<$0.1-0.2 pixels in the STIS data ({\it e.g.}
velocity zeropoint accuracies better than
0.6~km~s$^{-1}$ and 2.8~km~s$^{-1}$ for the E230M and G230M
spectra, respectively).  WAVECALs were obtained during the observations
and the  peakup procedure (ACQ/PEAK) was used for the acquisition
to center the source accurately, to about a 5\% of the slit width
(see www.stsci.edu/hst/stis/ for more details).
This accuracy was cross-checked for the
E230M spectra using as reference the narrow absorption features observed in
the Fe~{\sc ii}, Mg~{\sc ii} and Mg~{\sc i} resonance lines
produced by the interstellar medium.
The STIS spectra
were co-added after this check to improve the SNR of the data since
no variations are detected. The profiles have been
classified in Fig~1 by observing date. The left panel
displays the very asymmetric \ciii and \siii profiles
that led GdCV to conclude that these lines are
produced at the base of the wind. The \civ line observed $\sim 2$~hours
before is also displayed; the profile is
very similar to the \ciii and \siii lines.
Bluewards shifted
emission is also detected in  the \oxigen lines (see central panel)
observed in Feb. 2001. The \cii , \mgii
and \mgi lines are obtained in the same spectra.  A minimum is detected
at a similar velocity in the \mgii lines;
these lines have large optical depths in dense plasmas with $T\simeq 10^4$K.
A narrow absorption component, probably of interstellar origin,
is detected at the stellar rest velocity.
There also seems to be an absorption component in the
\mgi line at the same velocity. The last observations of the
\ciii and \siii lines are displayed in the right panel of Fig.1.
The profiles
changed significantly from 1993 to 2001, although the bluewards
shifted part of the profile remains stable.

\section{Interpretation and Analysis}

The detection of emission from high velocity gas
in \civ , \siii , \ciii ,  \oxigen , [O~{\sc I}] (Hamann 1994) indicates
that {\it there is a broad range of temperatures in the RY Tau
wind at similar velocities.} The wind properties seem to
be rather stable; {\it the emission peaks at similar
bluewards shifted velocities in observations obtained over 10 years},
compare Fig.~1 with Hamann (1994) that measures a blueshifted
[O~{\sc I}] emission peaking at -79 km/s.

Line ratios can be used to constrain the wind temperature
and density taking into account that the regions
traced by each spectral tracer only overlap partially.
Also occultation effects may be relevant (see Sect.4).
Line emissivities have been calculated for
a collisional plasma over a broad range of
densities ($n_e = 10^7 - 10^{13}$~cm$^{-3}$) and
temperatures ($\log T_e (K) = 4.0 -5.0$) using the
atomic parameters in the CHIANTI data base (Dere et al 1997)
and assuming solar abundances and ionization equilibrium.

\noindent
{\it The high temperature region:} from the
GHRS observations we derive: \civ/\siii~$=0.6 \pm 0.7$
and \siii/\ciii ~=$1.1 \pm 0.6$. Then from
Fig.~2 we derive: $\log (T_e) \simeq 4.8$ and
$ 9.5 < \log (n_e) <10.2$.
{\it Neither the line ratios nor the UV line fluxes
vary by more than a factor of 2 in the fifteen observations
available from
1979 to 2001 (GdCV)}. Notice that the high resolution data
indicate that the variable contribution from
0-velocity and redshifted material to the  \siii and  \ciii  line
fluxes does not  significantly change their line ratios
\footnote{
$\left[\frac {Si III]}{C III]}\right]_{GHRS} = 1.1 \pm 0.6$, $\left[\frac {Si III]}{C III]}\right]_{STIS} = 1.7 \pm 0.5$,
$\left[\frac {Si III]_{STIS}-Si III]_{GHRS}}
{C III]_{STIS}-C III]_{GHRS}}\right] = 4 \pm 3$
Thus the {\it average} physical conditions of this
hot plasma in the wind are similar to those
derived from the redshifted material.}.

\noindent
{\it The low temperature region:}
The  \cii emission feature consists of five spectral lines
(see Fig.~1 and Table~1). The individual features are not
resolved in RY~Tau; a broad blend is observed instead.
Some ratios among the individual lines in the feature
are {\it density sensitive} in the range
$n_e = 10^8-10^{11}$~cm$^{-3}$.
An accurate determination of the electron density in the
\cii emission region requires us to derive simultaneously the
ratios among the multiplet lines and the underlying
\cii profile. A least squares scheme has been used to solve
this inverse problem. The optimal fit gives
a high lower limit of $n_e > 10 ^{10}$~cm$^{-3}$
for the \cii emitting region, similar
to that determined from the \ciii line.
Note that the \cii profile is too
noisy to narrow this range (the contribution from
the Fe~II(uv3) line at vacuum wavelength 2328.11\AA\ is
negligible as shown by the weakness of nearby features
corresponding to the same multiplet).
The fit also indicates that the underlying \cii profile
is similar to the  \ciii (STIS) profile; {\it e.g.}
a blueshifted wind profile plus an excess from
0-velocity to the redwards shifted edge.
From that fit, we also obtain
$
\frac {F(C~{II]}_{2326.117})}{F(C~{ III]}_{1908.7})}
= 1.3 \pm 0.1
$.
This ratio is {\it temperature sensitive} and
from that we derive that $4.4 < \log (T_e) < 4.5$.

\section{Implications for wind models}

TTSs outflows are characterized by their collimation
at large scales, {\it i.e.}, it is expected that whatever
the ejection mechanism is, the wind terminal
speed corresponds to a longitudinal motion along
the collimation axis.  Current models make
use of magnetic fields to guarantee the large scale collimation.
In this section we use the models of G\'omez de Castro \&
Ferro-Font\'an 2005 (hereafter GdCFF) to compare to data.
Disk wind models by other authors have a similar kinematics
though the scales are different (see Uzdensky 2004 for a review).

In GdCFF models, the wind is launched from the disk corona
by a combination of thermal pressure and centrifugal force.
This results in  a radial expansion, away from
the rotation axis, at the base and produces an increasing
broadening of the profiles in this acceleration region.
After this, the restoring toroidal component of the
field narrows down the beam and the velocity field is
progressively dominated by the motion along the rotation
axis until it reaches the asymptotic regime, {\it i.e.}
flow along the axis at the terminal velocity and basically no
radial expansion. The evolution of the radial expansion,
$V_r$, with the height above the disk is plotted in the top
panel of Fig~3 for the warm disk winds model.
Line emission is produced at different locations in the flow
depending on densities and temperatures thus, the line profiles
depend on the region of the wind sampled
by each particular spectral indicator (as well as on
the inclination).

In order to compare these theoretical
predictions with the observations, we have defined two observables
that are measured only from the bluewards shifted emission of the relevant
spectral tracers  to avoid the pollution by the contribution
of the accretion flow. Notice that line
emission from high temperature infalling plasma is produced at scales
$<<R_*$ pumped by the X-ray radiation released at the accretion
shocks. Thus occultation by the stellar disk causes the
contribution to the line flux from accretion to be redwards
shifted (see Beristain et al (2001) for more details).
These observables are the velocity
dispersion, $dV$, and the line centroid,$V_c$. The line
profiles may be understood as
velocity-flux plots (or $V-f$ plots). The centroid of the
bluewards shifted region of the profile  is defined as,
$
V_c = \frac {\Sigma _{i=1}^{i=N}f_i (V_0+i\Delta V)}{\Sigma _{i=1}^{i=N}f_i}
$
and the dispersion, $dV$, as,
$
dV = \sqrt{\frac{\Sigma _{i=1}^{i=N}f_i ((V_0+i\Delta V)^2 - V_c^2)}
{\Sigma _{i=1}^{i=N}f_i}}
$
for a uniformly sampled profile with step $\Delta V$ and $N$ points.
$V_0$ represents the velocity at the blue edge of the profile
and $N = ABS(V_0/\Delta V)+1$. These observables have been
calculated for  three spectral tracers that sample very different
plasma densities: \siii, \ciii and \oxigen.
The
critical electron densities for the collisional quenching of
the \ciii and \oxigen lines are
$1.5 \times 10^{10}$~cm$^{-3}$ and $ 0.5
\times 10^7$~cm$^{-3}$ respectively. For comparison,
the critical density of the  \siii line is
$2.0 \times 10^{12}$~cm$^{-3}$ for a fiducial temperature
of $2.5\times 10^4$K.
$V_c$ and $dV$ are plotted in the bottom panel of Fig 3 for the
GdCFF theoretical predictions and various inclinations;
notice that the \oxigen line is predicted to be formed in the
rarified expanding region while the \ciii and \siii lines are
predicted to be formed in the hot disk corona where the
wind is accelerated.

$V_c$ and $dV$ have been computed for the \oxigen , \ciii and \siii
profiles observed in RY~Tau. To take into account the effect of the noise
on these observables, we have computed these values both on the
observed profiles and on filtered profiles obtained by applying
a white noise filter to the data. As shown in Fig.~3, the major effect
of the noise is to increase the velocity dispersion and shift
the centroid some few km~s$^{-1}$ to the blue; however the
overall trend remains. Two important results are
derived from the figure:
\begin{enumerate}
\item The line centroid shifts to higher velocity
as the density traced by the spectral line
becomes lower as expected (see critical densities),
\item The velocity dispersion is high,
$> 40$km~s$^{-1}$, and similar for all the tracers.
In addition, the profiles are very asymmetric (see Fig.~1).
Rather constant velocity dispersions  and asymmetric
profiles are expected for very small inclinations,
while for large inclinations the profiles should be
symmetric and the dispersion increase steadily with
the line centroid. However, all evidence points to
RY~Tau being observed cloes to edge-on (see Sect.~1).
\end{enumerate}
Thus RY~Tau observations cannot be
reconciled with models
based on the centrifugal launching of the TTSs jets
from the accretion disk.

Another important piece of evidence comes from the
emitting volumes. The volume of plasma producing the
observed \siii and \ciii flux can be derived from the
line emissivity\footnote{The
emissivities are: j(\civ ) = $1.44 \times
10^{-2}$erg~cm$^{-3}$~s$^{-1}$ for $\log (T_e) =4.8$ and $\log (n_e) =11$;
j(\siii ) = $1.49 \times
10^{-2}$erg~cm$^{-3}$~s$^{-1}$ for $\log (T_e) =4.8$ and $\log (n_e) =11$;
and j(\ciii ) = $0.97 \times
10^{-3}$erg~cm$^{-3}$~s$^{-1}$ for $\log (T_e) =4.8$ and $\log (n_e) =11$.}.
The emitting volume producing the \siii emission is
$2.07\times 10^{31}$cm$^3$ ({\it i.e.} (0.24R$_{\odot}$)$^3$)
and similar to the derived for the \ciii and \civ lines. Thus,
line emission is produced in either a small structure
or in dense clumps in the outflow. Notice that even
adopting a filling factor of 0.1\%--1\% for the clumps, as that
derived from observations of optical jets (Liseau et al 1996), the volume
of the region producing the observed \ciii and \siii profiles
just increases to about $ 1.1-2.4$~R$_{\odot}$.
Thus, it is highly likely that mass ejection occurs from
the stellar surface.

As RY~Tau is a fast rotator the
centrifugal lever arm from the stellar surface is an efficient
source of thrust (see {\it e.g.} Sakurai 1987).
Thus, the wind could be
ejected from the stellar surface  from
open field structures like those observed in the
Solar corona. The interaction between the wind and
the circumstellar material (and a probable slow wind)
could produce shocked knots like the Corotating
Interaction Regions (CIRs) observed in the interplanetary
medium and in young stars like AB~Dor (G\'omez de
Castro 2002) that could contribute to the UX~Ori
phenomenon observed in RY~Tau.

This suggestive possibility would indicate that
the observed profiles represent just an average
of many small scale outflows and although these
would not be stationary on a one-by-one basis, the
average of all of them could produce a rather
stable profile (provided that the spectral resolution
is not good enough to resolve the individual components).
If the magnetic field configuration is not axisymmetric
(as observed in cool main sequence stars) this wind would
naturally lead to the formation of nonaxisymmetric outflows.
Tracking the variability of the
\civ and \ciii profiles is crucial for testing
this possibility.

\noindent {\bf Acknowledgments:} We thank an anonymous referee for
helping us to sharpen the definition of the observables in Fig.~3.

\clearpage

\begin{figure}
\includegraphics[width=12cm]{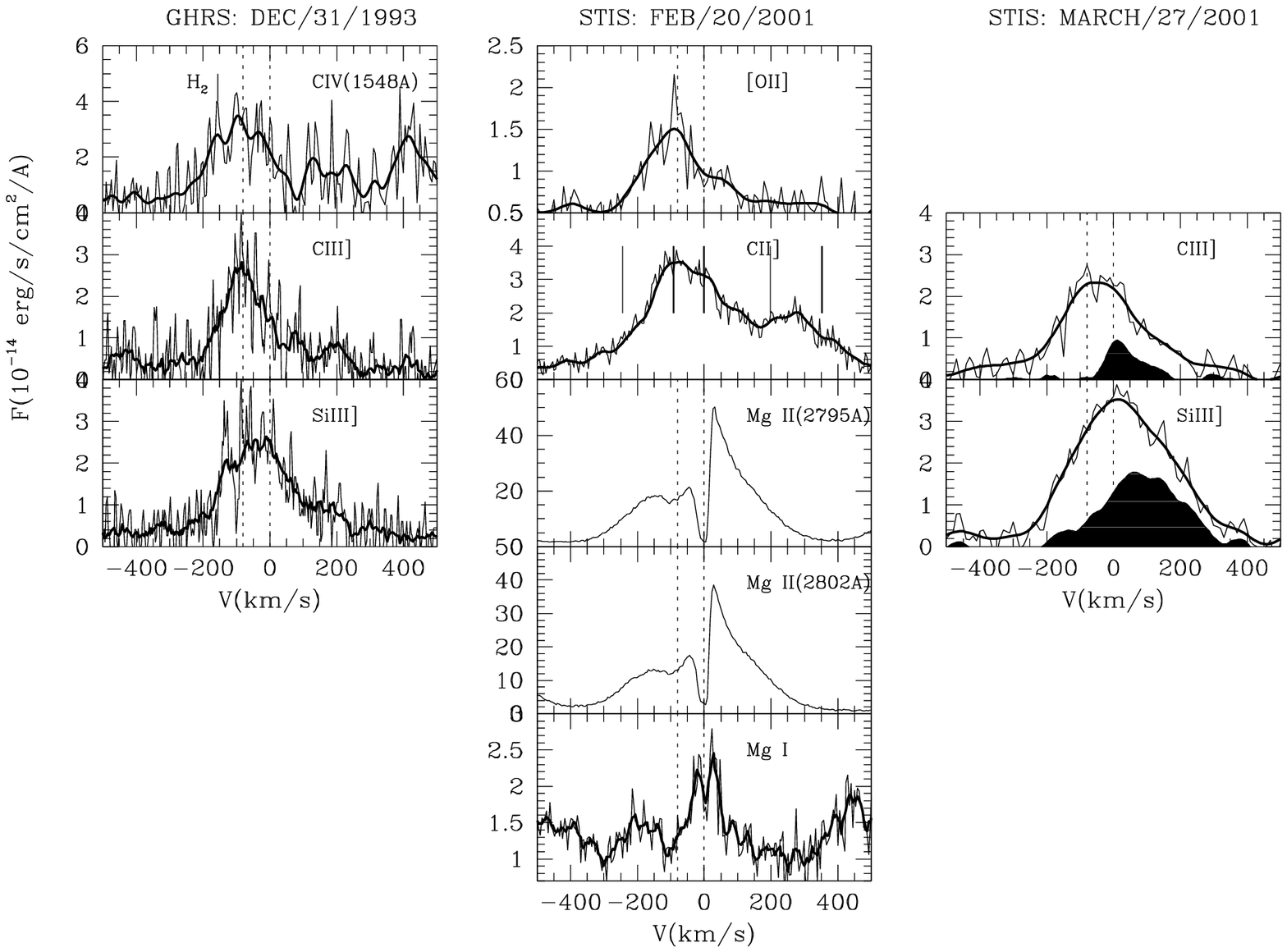}
\caption{ UV lines observed in RY~Tau
and used in this work; RSDP processed data (see Table~1 for the
binning) are plotted with a thin line and the 3-pixels average
profile with a thick line. The rest wavelength of the lines and
the velocity of the unresolved jet at $\simeq -80$ km/s (from GdCV
and Hamann 1994) are marked with dashed lines. {\it Left panel}:
Observations obtained in Dec. 31st, 1993 with the GHRS. The \civ
line is blended with a very weak H$_2$ feature (see Ardila et al
2002). {\it Central panel}: Profiles from STIS observations
obtained in Feb. 20-21, 2001. The 5 components of the \cii
multiplet are marked. {\it Right panel}: \ciii and \siii lines
observed in March 27th, 2001. Both lines show an excess of flux in
the red wing compared with the 1993 observations; this excess is
shaded in the figure.}
\end{figure}

\clearpage
\begin{figure}
\includegraphics[width=12cm]{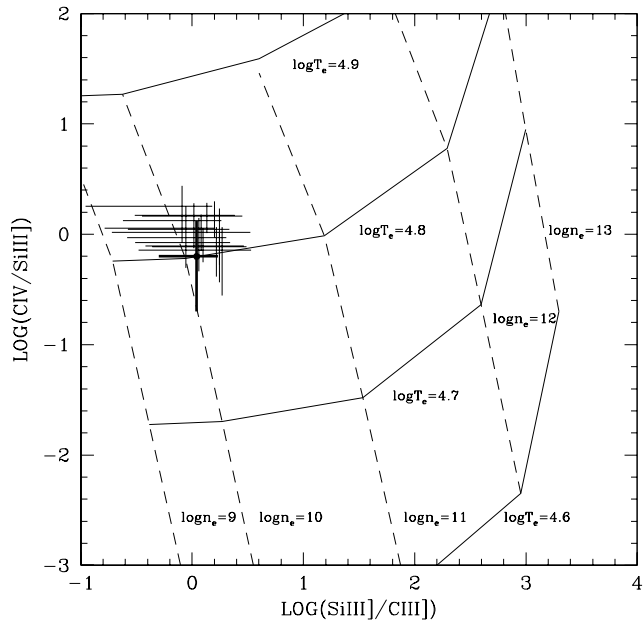}
\caption{ \civ/ \siii versus  \siii/ \ciii flux
ratios (in erg/s/cm$^{-2}$) for collisional plasmas with  electron
densities between $10^9-10^{13}$cm$^{-3}$, temperatures
$\log(T_e)$=4-5 and solar abundances. The value (with the error bars) derived from
high resolution observations is marked with thick line;
the 13 low resolution IUE observations are plotted with thin lines.
Isothermal and isochoric lines are plotted with
continuous and dashed lines respectively.
}
\end{figure}

\clearpage

\begin{figure}
\includegraphics[width=12cm]{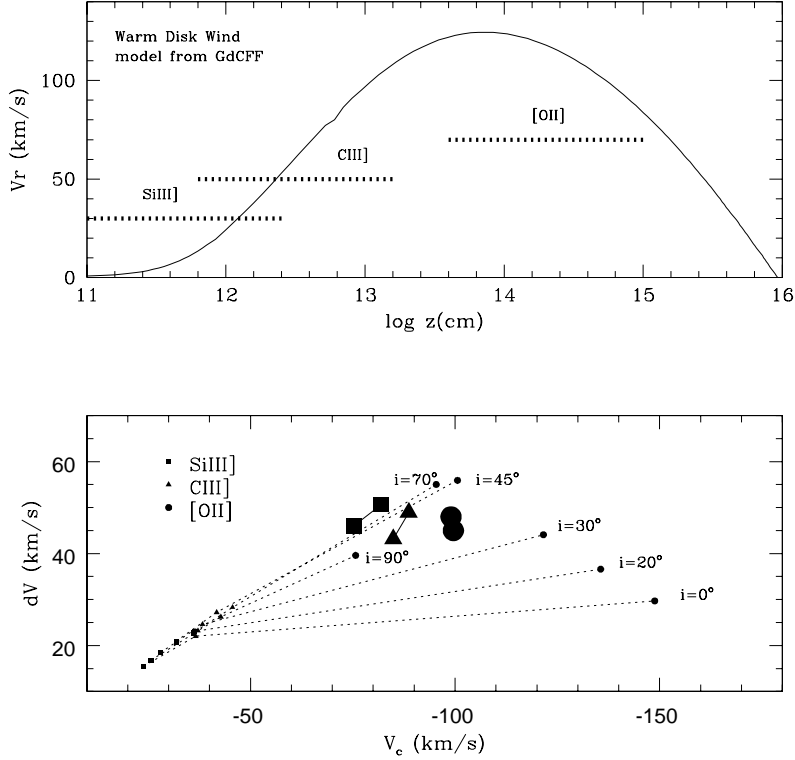}
\caption{{\it Top:}  variation of the wind expansion
velocity with respect to the jet axis, $V_r$, with the height above
the disk, $z$, from GdCFF model. The location of the \siii, \ciii and
\oxigen emission regions is sketched (see GdCFF for the detailed
calculations). {\it Bottom:} Comparison between the theoretical
predictions from the GdCFF model and RY Tau observations.
$V_c$ and $dV$ values measured from
the \siii, \ciii and \oxigen profiles observed in RY Tau are marked with
big squares, triangles and circles respectively.
$V_c$ and
$dV$ are plotted from the \siii, \ciii and \oxigen profiles derived
from the theory (small symbols) for various inclinations.  High $dV$ values
are obtained from the observed profile while the slightly
smaller $dV$ values are measured over filtered profiles (see text). }
\end{figure}

\clearpage

\setlength{\oddsidemargin}{-2cm}
\begin{table}
\begin{scriptsize}
\caption{HST/STIS observations.}
\begin{tabular}{lllllcccl}
\hline
Inst. & Grating & Res. &Scale$^{(a)}$& Slit & Spec. Range & Obs. Date  & Target$^{(b)}$           & Flux $^{(c)}$ \\
           &         &      &         &     &             &            & Features &                     \\
           &         &      &\AA /elem &    & \AA         &(dd-mm-yyyy)&                  &$10^{-14}$ (erg/s/cm$^2$) \\
\hline
GHRS$^{(d)}$  & G160M & 20000 &0.069& LSA$^{(e)}$& 1532 - 1567 & 31-12-1993& C~{\sc IV} & F(C~{\sc IV})$=8.2\pm 7.1^{(g)}$\\
     & G200M & 20000 & 0.078 &LSA&1880 - 1920 & 31-12-1993& Si~{\sc III}], C~{\sc III}] & F(Si~{\sc III}])$=13\pm 4$\\
     &      & &       &             &           &                             & F(C~{\sc III}])$=12\pm 3$ \\
STIS$^{(g)}$ & E230M & 30000 &1/60,000&0".2x0".2&  2303 - 3111 & 19-02-2001 &  [O~{\sc II}], [C~{\sc II}],Mg~{\sc II}, Mg~{\sc I} & F([O~{\sc II}])$=5.5\pm 1.4 $  \\
     & E230M & 30000 &1/60,000& 0".2x0".2&2303 - 3111 & 20-02-2001 &[O~{\sc II}], [C~{\sc II}], Mg~{\sc II}, Mg~{\sc I} &
F([C~{\sc II}]$_{2326.1}$)$=19\pm 2$ \\
     & E230M & 30000 &1/60,000& 0".2x0".2&2303 - 3111 & 20-02-2001 &[O~{\sc II}], [C~{\sc II}], Mg~{\sc II}, Mg~{\sc I} &\\
     & G230M & 10500 &0.09 & 52"x0".2&1839 - 1929 & 27-03-2001 &Si~{\sc III}], C~{\sc III}] &  F(Si~{\sc III}])$=26\pm 3$\\
 & &       &       &           &  &           &                             & F(C~{\sc III}])$=15\pm 3$ \\
\hline
\end{tabular}
\end{scriptsize}
\begin{scriptsize}
\begin{tabular}{ll}
$^{(a)}$ & The element is a diode for GHRS spectra and a pixel for STIS spectra.\\
$^{(b)}$ & The vacuum wavelengths of the target features are:
C~{\sc IV}(1548\AA , 1550\AA), Si~{\sc III}](1892\AA), C~{\sc III}](1908\AA), [O~{\sc II}] (2471\AA), \\
& [C~{\sc II}](2324.2\AA,2325.4\AA,2326.1\AA,2327.6\AA,2328.8\AA),
Mg~{\sc II} (2796\AA,2803\AA),Mg~{\sc I} (2853\AA) \\

$^{(c)}$ &The Interstellar reddening law has been used since the extinction
law towards T Tauri systems is not known.\\
    & Parameters: A$_V$=0.55mag; A$_{CIV}$/A$_V$ = 2.60;
A$_{SiIII}$/A$_V$ = 2.66; A$_{CIII}$/A$_V$ = 2.66; A$_{CII}$/A$_V$ = 2.60;
A$_{OII}$/A$_V$ = 2.44.\\
& The flux of the [C~{\sc II}](2326.1\AA) has been derived from the fitting of the \cii multiplet (see text).\\
$^{(d)}$ & Goddard High Resolution Spectrograph removed in February 1997\\
$^{(e)}$ & Large Science Aperture (2'')\\
$^{(f)}$ & The H$_2$ (3-8 R(3)) line contribution to the C~{\sc IV} flux is
negligible (see Ardila et al 2002).\\
$^{(g)}$ & Space Telescope Imaging Spectrograph stopped science operations in August 2004\\
\end{tabular}
\end{scriptsize}
\end{table}


\begin{thebibliography}{}

\bibitem{}
Ardila, D. R., Basri, G., Walter, F. M., Valenti, J. A.,
Johns-Krull, C. M., 2002, ApJ, 566, 1100
\bibitem{}
Beristain, G. Edwards, S., Kwan, J., 2001, ApJ, 551, 1037
\bibitem{}
Dere, K. P., Landi, E., Mason, H. E., Monsignori Fossi, B. C., Young, P. R.,
1997, A\&AS, 125, 149
\bibitem{}
Dupre\'e, A. K., Brickhouse, N. S., Smith, Graeme H., Strader, Jay, 2005, ApJ, 626, L59
\bibitem{}
G\'omez de Castro, A.I., Verdugo, E., 2001, ApJ, 548, 976
\bibitem{}
G\'omez de Castro, A.I., 2002, MNRAS, 332, 409
\bibitem{}
G\'omez de Castro, A.I., Ferro-Font\'an, C., 2005, MNRAS, 365, 569
\bibitem{}
Grinin, V.P., 1992, A\&AT, 3, 17
\bibitem{}
Hamman, F., 1994, ApJS, 93, 485
\bibitem{}
Hartmann, L., Stauffer, J.R., 1989, AJ, 97, 873
\bibitem{}
Liseau, R., Huldtgren, M., Fridlund, C.V.M., Cameron, M., A\& A, 306, 255
\bibitem{}
Sakurai, T., 1985, A\& A, 152, 121
\bibitem{}
Uzdensky, D., 2004, Ap\&SS, 292, 573
\end{thebibliography}
\end{document}